\documentclass[12pt]{article}
\usepackage{graphics}
\usepackage{epsfig}
\usepackage{graphicx}
\usepackage{graphicx}
\newcommand{\ve}{\varepsilon}

\renewcommand\thesubsubsection{\arabic{subsubsection}}

\begin{document}

\begin{center}
{\large\bf Absorption tomography of laser induced plasmas
 with a large aperture} 

\vskip 1cm
{\bf S.~V.~Shabanov${}^{1,2}$ and 
I.~B.~Gornushkin${}^2$}

\vskip 1cm
${}^1$ {\it Department of Mathematics, University 
of Florida, Gainesville, FL 32611, USA}

${}^2$ {\it BAM Federal Institute for Materials Research and Testing, Richard-Willst\"atter-Strasse 11, 12489 Berlin, Germany}

\end{center}

\begin{abstract} An emission tomography of laser-induced plasmas
employed in the laser induced breakdown spectroscopy (LIBS)
requires long signal integration times during which
the plasma cannot be considered stationary.
To reduce the integration time,
it is proposed to measure a plasma absorption in parallel rays 
with an aperture that collects light
coming from large fractions of the plasma plume
at each aperture position.
The needed spatial resolution is achieved by a special numerical data
processing. 
Another advantage of the proposed procedure is 
that inexpensive linear CCD or non-discrete (PMT, photodiode)
detectors can be used instead of costly 2-dimensional detectors.

\end{abstract}

\subsubsection {Large apertures and the Abel inversion} 
It is assumed that a plasma
plume created by a laser ablation is axially symmetric 
(the symmetry axis coincides with the laser ray). 
Let the coordinate system be set so that the symmetry axis
is the $z-$ axis. Then the emissivity is a function 
$\varepsilon=\varepsilon(r,z,t,\nu)$ where $r=(y^2+x^2)^{1/2}$,
$t$ is the time, and $\nu$ is the frequency. A measured quantity is
the intensity $I(y,z,t,\nu)$ of light
per unit time and unit frequency
along the rays 
through an infinitesimal area element $\Delta A=\Delta y\Delta z$
centered at the point $(0,y,z)$ that are in a narrow solid angle 
$\Delta \Omega$ about the line parallel to the $x-$axis:
\begin{equation}
\label{1}
I(y,z,t,\nu) = \Delta A\Delta\Omega \int_{-\infty}^\infty
dx\,\varepsilon\Bigl(\sqrt{y^2+x^2}, z,t,\nu\Bigr)=2\Delta A\Delta\Omega
\int_y^\infty \frac {dr\, r\, \ve(r,z,t,\nu)}{\sqrt{r^2-y^2}}\,.
\end{equation}
It is well-known that Eq.(\ref{1}) can be solved
for $\ve(r,z,t)$ by the Abel inversion \cite{1,2}. A plasma plume
has a finite size that determines a cut-off of the infinite
integration limits in (\ref{1}). 
Equation (\ref{1}) is valid only for infinitesimal $\Delta A$
and $\Delta\Omega$ and so is the approximation of the parallel
rays that allows for the subsequent Abel inversion. 
In typical LIBS experiments, the smallness of $\Delta A$  
is  provided by a narrow spectrometer slit and small pixel
size of a detector, while 
$\Delta \Omega$ is small due to a low acceptance angle
of a spectrometer (high $f-$numbers). 
To compensate
for small factors $\Delta A$ and $\Delta\Omega$, 
a long integration time $T$ is used
to collect enough energy $E=\int_0^T dtI(y,z,t,\nu)$
of photons emitted by the plasma along parallel rays.
A typical integration time in LIBS plasma experiments
is $T\sim 1\, \mu s$ or even higher.
If the plasma is not stationary during the time $T$,
the collected data correspond to a time-averaged intensity
and the reconstructed emissivity may not be accurate.
This is especially relevant for studying elemental 
contents of plasmas expanding into 
an ambient gas because of percolation processes 
at a rapidly moving plasma-gas interface.   
Numerical simulations of LIBS plasma dynamics
show that this is often the case, especially for earlier
stages of the plasma evolution (see, e.g., \cite{3,4}).

Here it is proposed to increase an aperture through which
parallel rays are collected in order to reduce the integration
time so that the observed intensity is:
\begin{equation}
\label{2}
I_E(y) = \Delta z\Delta\Omega \int_{y-\Delta}^{y+\Delta}
du \int_{-\infty}^{\infty} dx\,
\varepsilon\Bigl(\sqrt{u^2+x^2}\Bigr)\equiv
\int_{y-\Delta}^{y+\Delta}
du I(u)\,.
\end{equation}
where the dependence on $z$, $t$, and $\nu$ is now suppressed as 
it is irrelevant for the discussion. Equation (\ref{2})
states that all photons traveling {\it parallel} to 
the $x-$axis and coming
through a rectangular aperture of an infinitesimal
height $\Delta z$ and finite
width $2\Delta$ and centered at a distance $y$ from 
the $x-$axis are collected. An experimental setup to which
Eq. (\ref{2}) applies is discussed below (see Fig.~\ref{fig1}).
In the limit $\Delta = \Delta y/2
\rightarrow 0$, Eq. (2) turns into Eq.(\ref{1}).
So, the signal can be amplified  
roughly by the factor of $2\Delta/\Delta y$.
A spatial resolution $\Delta y$ in a typical
experimental setup \cite{5,6,7} is determined by 
a pixel size of a camera
used to register $I(y)$, $\Delta y \sim 0.02\, mm$.
If the aperture width is taken to be of a typical size of a LIBS plasma
plume, $2\Delta \sim 2\, mm$, up to two orders in magnitude of
the signal strength can be gained.
How is then the spatial resolution restored?
It is proved below that, given a function 
$I_E(y)$, the function $I(y)$ can be {\it uniquely} determined
from (\ref{2}) for {\it any} $\Delta$, 
provided $I(y)$ satisfies the condition 
that $I(y)$ vanishes for $|y|>D$ for some $D$, which is always 
the case for a physical $I(y)$ because a plasma plume has a finite
size. 
Thus, the proposed procedure entails:\\ \\ 
{\it (i)} Taking intensity measurements
of $I_E(y)$ at a set of positions $y=y_n$;\\ 
{\it (ii)} Reconstructing
the intensities $I(y_n)$ from the data $I_E(y_n)$;\\ {\it (iii)}
The Abel inversion
for $I(y_n)$. 

A key issue for the data collection {\it (i)} is
to separate parallel and angled rays passing through the aperture.
Note that for a non-infinitesimal
$\Delta$ there should be rays coming through the aperture
at a finite angle to its normal and, hence, $\Delta \Omega$
can no longer be considered infinitesimal, thus, invalidating
the approximation of parallel rays in (\ref{2}). 
This is indeed true if the plasma {\it emission}
is collected by the slit. The problem can be avoided if
the collimated light {\it absorbed} by the plasma is measured. An experimental
setup to achieve this goal is presented in Section 2 
(see Fig.~\ref{fig1}).

An explicit algorithm to carry out {\it (ii)} is given
in Section 3. The spatial resolution is determined by 
the difference $\Delta_s= y_n-y_{n-1}$ in positions of the center 
of the aperture. Numerical simulations with synthetic data representative
for LIBS plasmas are presented in Section 4.
There is a variety of numerical methods available to carry out 
the Abel inversion (see, e.g., \cite{abelinversion},
and more recent applications to LIBS plasmas \cite{6,7}). 
So the part {\it (iii)} will not be discussed here.

\subsubsection{Plasma absorption experiments} 

\begin{figure}
 \centering
\includegraphics[height=6cm,width=12cm]{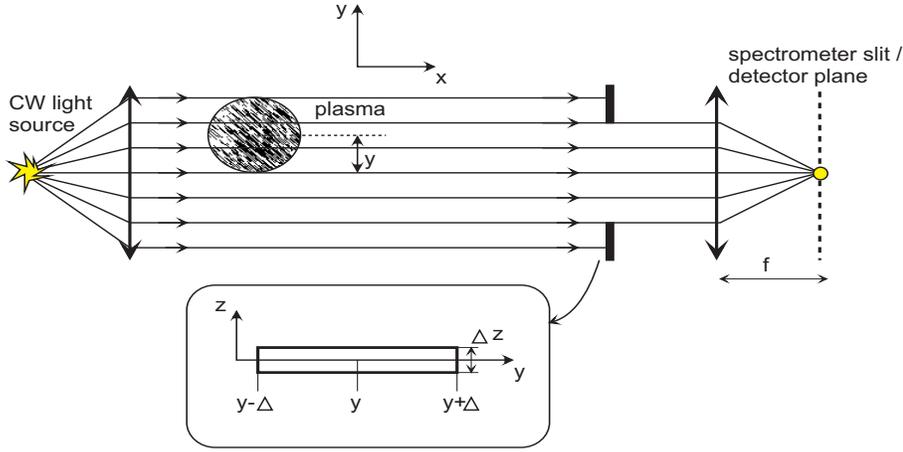}
\caption{\small An experimental setup for plasma absorption measurements
with a large aperture. A CW light source is in focus of a lens
to create a flux of parallel rays. A plasma plume with the symmetry
axis normal to the figure plane is illuminated by the parallel rays.
The parallel flux across a rectangular aperture is focused onto
a detector (a spectrometer slit or CCD camera). The aperture plane is
normal to the flux. It is positioned at a distance large
enough to neglect contributions of the plasma plume emission.
The aperture geometry is depicted in the inset at the bottom of the 
figure. The height $\Delta z$ is small enough to neglect variations
of the plasma absorption along the $z-$direction. The aperture 
width $2\Delta$ can be of the size of the plasma plume or larger.
The intensity in 
parallel rays is measured for different values of $y$, the distance
between the plasma symmetry axis and the optical axis of the system.
Measurements at a fixed value of $y$ have no spatial resolution
by themselves 
if $\Delta$ is large. The spatial resolution is {\it restored} by 
a numerical processing of data collected at different values of $y$.}
\label{fig1}
\end{figure}

The proposed hypothetical experimental setup
is shown in Fig.~\ref{fig1}. 
A point light source 
is placed in a focus of lens so that after the lens
the light propagates
parallel to its optical axis. The light goes through 
a plasma plume. Some of the photons will be absorbed by 
the plasma and then re-emitted. However the re-emission occurs
in the whole solid angle of $4\pi$, i.e., the absorbed
light no longer contributes to the parallel flux. Since the intensity 
of the plasma radiation is inversely proportional to 
the squared distance from the plasma center, by placing
the aperture at a distance large enough from the plasma center
the contribution of the plasma radiation through the aperture
becomes negligible as compared to that of the parallel light.  
The plasma emissivity, $\ve(r)$, is determined by the 
absorption coefficient $\kappa(r)$
and the black body radiation function $B(r)$, $\ve(r)=\kappa(r)B(r)$,
provided, of course, that the plasma is at a local thermodynamic
equilibrium (see, e.g., \cite{zel}).
So, by measuring the difference between the fluxes in parallel rays
piercing through the aperture area $\Delta A= 2\Delta z\Delta$
with and without the plasma plume inserted, the contribution 
of the plasma emissivity 
expressed as the attenuation of the parallel flux across $\Delta A$
is obtained. The plasma plume symmetry axis is positioned
at a distance $y$ from 
the optical axis of the right lens (the lens behind the slit). 
The center of the aperture
also lies on this optical axis. The 
parallel rays that came through the aperture are focused on 
the spectrometer slit. In this setting, the spectrometer-detector
combination is only used 
to achieve the spectral resolution because all the 
light that came through the aperture of width $2\Delta$ 
is focused to a {\it single}
point of the spectrometer slit. 
Thus, there is no spatial resolution
in measurements at a {\it fixed} value of $y$.
An analogy to the data collection in
the aforementioned {\it emission} experiments,
e.g. \cite{6,7}, would correspond to the case when
the right lens is removed in Fig.~\ref{fig1} so that
the light is collected by {\it all} pixels of a detector
evenly distributed along the spectrometer slit and both
the spatial and spectral resolutions are achieved. 
In contrast, here all the light is  focused
onto a {\it single} pixel by the right lens in Fig.~\ref{fig1}. 
Thus, the signal amplification 
is achieved at the price of loosing the spatial resolution
in measurements. However,
as noted before, the spatial resolution is restored
by a numerical data processing. 
It is noteworthy that 2D CCD detectors are no longer required
to spatially resolve the plasma radiation. Inexpensive linear
detectors (if the spectral dimension is to be retained) or
photomultiplier tubes (PMT) can be used instead.

The measurements of $I_E(y)$ are taken at different positions
of the center of the aperture $y=y_n=n\Delta_s$, $n=0,\pm 1,
\pm 2,...,\pm N$, where $\Delta_s$
is the slit center displacement relative to the plasma plume
symmetry axis (it is more convenient to create plasma
plumes at different positions $y_n$). The step $\Delta_s$ 
can be made as small 
as desired (or possible to achieve). The limiting positions 
$y=\pm N\Delta_s$ are chosen so that  
the studied plasma plume absorption 
does not  contribute to the parallel flux through the aperture
(i.e., $N\Delta_s-\Delta$
exceeds the plasma plume radius).

The light source for absorption measurements 
must be bright enough, i.e., comparable
to the plasma own spectral emission, otherwise
the gain by the large aperture is rendered useless by 
a flux weaker than that of the plasma emission. Possible solutions
of this problem are as follows. First, another laser-induced
plasma plume can be used as the point light source. It is 
just as bright as the studied plasma plume, has a broadband 
spectrum, and a short pulse duration suitable 
for time-resolved measurements.
Plasma plumes as the light source for absorption
measurements have been used in experiments reported in \cite{g1}.
Second, super-continuum white light lasers \cite{laser} with 
a broadband output spectrum can be used as a light source
for absorption measurements. Finally,
if the spectral resolution is not relevant,
a single frequency laser can be used to study plasma emissivity 
at a particular frequency that coincides with 
one of the plasma emissivity spectral lines \cite{g2}. 
The problem of an insufficient spectral brightness is also resolved
in this case.
 
\subsubsection{The reconstruction algorithm} 
 
The integral equation (\ref{2}) is a linear equation.
So its solution is the sum of a general solution $I_0(y)$ of the 
homogeneous equation (when $I_E=0$) and a particular solution of 
the non-homogeneous equation (\ref{2}). It is assumed
that $I_E(y)$ cannot be a non-vanishing constant (a plasma 
plume has a finite size). Differentiating Eq. (\ref{2})
with respect to $y$ and shifting the argument $y\rightarrow y-\Delta$,
one infers:
\begin{equation}
\label{3}
I(y) = I(y-2\Delta) + I'_E(y-\Delta)\,.
\end{equation}
If $I_E=0$, then $I(y)=I_0(y)=I_0(y-2\Delta)$ 
is a periodic function that is 
a linear combination 
of $\exp(i\pi ky/\Delta)$ for $k=\pm 1,\pm 2,...$
(the case $k=0$ is not possible for $I_E=0$ in
Eq. (\ref{2})).
By physical boundary conditions, $I_E(y)=I(y)=0$ for 
all $|y|>D$ and some $D$. No periodic $I_0(y)$ satisfies
this condition. Thus, under this boundary condition, $I(y)$ is 
uniquely recovered from $I_E(y)$. The solution can be 
written in the form:
\begin{equation}
\label{4} 
I(y) = \frac 12 \sum_{q=0}^\infty \Bigl\{
I'_E\Bigl(y-\Delta(2q+1)\Bigr) - 
I_E'\Bigl(y+\Delta(2q+1)\Bigr)\Bigr\}\,.
\end{equation}
It can be verified by its substitution into the right side
of Eq. (\ref{2}). To understand the structure of the solution
(\ref{4}), it might be instructive to work out a simple 
analytic example of reconstructing $I(y)=A=const$ for
$|y|\leq \Delta$ and $I(y)=0$ otherwise. The function $I_E(y)$
is easy to compute by Eq. (\ref{2}),
$I_E(y)=A (2\Delta - |y|)$ if $|y|\leq 2\Delta$ and $I_E(y)=0$
otherwise. Then $I_E(y)$ can be 
substituted into the right side of Eq. (\ref{4}) to see how
$I(y)$ is recovered through cancellations in the sum.
In this simple case, 
the function $I(y)$ is reconstructed in the interval $|y|\leq \Delta$
by the first
term $q=0$ in the sum (\ref{4}). The other terms vanish
for $|y|\leq \Delta$ and are needed
to make $I(y)=0$ for $|y|>\Delta$. So, for a plasma plume 
of a radius $R$, only first $R/\Delta$ terms are needed to
reconstruct $I(y)$ in the interval $|y|\leq R$.

Put $\Delta = m\Delta_s$ for 
some integer $m$. Suppose that the data function $I_E(y)$
is taken at the grid points $y_n = n\Delta_s$ where
$n=-N,-N+1,...,N-1,N$ so that $I_E(y)=0$ for all $y \leq y_{-N}$
and $y\geq y_N$. Consequently, this implies that  
$I(y) = 0$ for all $y\leq y_{-N+m}$ and $y\geq y_{N-m}$.
The number $N$ is chosen so that 
$(N-m)\Delta_s$ exceeds the plasma plume radius.
In practice, the data collection
should simply start at $|y|$ large enough to see no signal from
the plasma.  Then $y$ is changed with the step
$\Delta_s$ until the signal vanishes again. Equation (\ref{3})
becomes the recurrence relation:
\begin{equation}
\label{slit3}
I(y_n) = I(y_{n-2m}) + I'_E(y_{n-m})\,.
\end{equation}
With the boundary conditions imposed on the data set, 
it follows from
the first $2m$ relations in (\ref{slit3}) that
\begin{equation}
\label{slit4}
I(y_n) = I'_E(y_{n-m})\,,\quad n=-N+m+1,...,-N+3m\,.
\end{equation}
The values $I(y_n)$ for $n>-N+3m$ are then determined 
recursively by 
(\ref{slit3}) because the preceding  values $I(y_{n-2m})$
needed to initiate the recurrence relation (\ref{slit3})
are now known from (\ref{slit4}). The remaining problem
is  to calculate $I_E'(y_n)$ from the set $I_E(y_n)$.

Let $\tilde{I}_E(k)=\int dy I_E(y)e^{iky}$ 
be the Fourier transform of $I_E(y)$.
The Fourier transform of $I_E'(y)$ is 
$ik\tilde{I}_E(k)$. Thus the  values
$I'_E(y_n)$ can be found by a discrete Fourier transform:
\begin{equation}
\label{ff}
I_E(y_n) \stackrel{F}{\longrightarrow} \tilde{I}_E(k_n)
\longrightarrow ik_n\tilde{I}_E(k_n) 
\stackrel{F^{-1}}{\longrightarrow} I'_E(y_n)\,,
\end{equation}
where $F$ implies taking the fast Fourier transform of the data
set $I_E(y_n)$ to obtain the set $\tilde{I}_E(k_n)$, 
and $F^{-1}$ is the inverse of $F$.
For analytic functions, the accuracy of a numerical differentiation
by the spectral (Fourier) method is superior
to any finite differencing method \cite{fron}. The spatial 
resolution of the reconstructed intensities $I(y_n)$ is 
determined by $\Delta_s$. It could even be better than
in conventional settings where spectrometers with a pixel
detector are used 
for the spatial resolution, provided it is possible 
to achieve $\Delta_s$
smaller than a spectrometer pixel size. 

\subsubsection{Numerical simulations with synthetic data}

\begin{figure}
 \centering
\includegraphics[height=6cm,width=17cm]{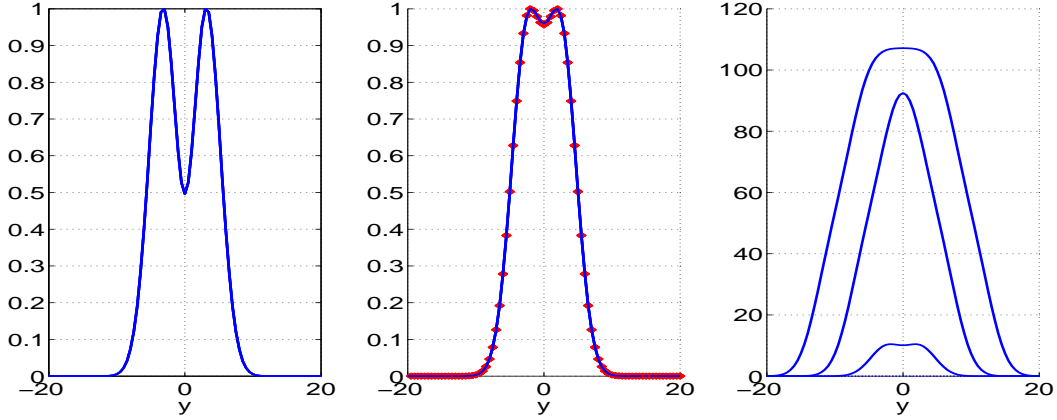}
\caption{\small {\sf Left panel:} The emissivity as a 
function of $r=|y|$ (the horizontal axis) 
as described in the text (arbitrary units). The plasma 
radius is roughly $R=10$.
{\sf Middle panel:} The intensity in parallel rays $I(y)$
calculated by Eq. (\ref{1}) with the factor $\Delta A\Delta\Omega$
omitted. It is normalized so that its maximal value is one.
The (red) dots indicate the values $I(y_n)$ reconstructed
from $I_E(y_n)$ for $\Delta=R$ 
by the data processing algorithm of Section 3.
{\sf Right panel:} The intensities $I_E(y)$ for
three different aperture widths. 
The upper, middle, and lower curves are the graphs
of $I_E(y)$ computed by Eq. (\ref{2}) with $\Delta = R$,
$\Delta = R/2$, and $\Delta = R/10$, respectively. 
} 
\label{fig2}
\end{figure}
The emissivity is taken in the form $\ve(r)=
\ve_0 \exp(-r^2/2\sigma^2)(a - \cos(br))$ which 
resembles the emissivity profile typically observed
in LIBS plasmas \cite{7}. The parameters 
are chosen as 
$\ve_0=1$, $\sigma=3$, $a=1.4$, and $b=0.5$ (arbitrary units).
The graph of $\ve(r)$ is shown in the left panel of 
Fig.~\ref{fig2}. As seen from the figure,
the plasma plume radius can be set as
$R=10$ in these units.
The intensity $I(y)$ in parallel rays is calculated 
by means of Eq. (\ref{1}) (the geometrical factor 
$\Delta A\Delta\Omega$ is omitted). It is shown
as the solid (blue) curve in the middle panel 
of Fig.~\ref{fig2}.
The function $I_E(y)$ calculated by means of Eq. (\ref{2})
is shown in the right panel of Fig. (\ref{2}).
The top curve corresponds to the aperture width $2\Delta=2R$.
For the middle
curve, the aperture width is reduced twice $\Delta = R/2$, 
and the bottom
curve corresponds to $\Delta = R/10$. 
In the conventional experimental settings,  
about $10$ to $20$ measurement across the plasma diameter
are taken. So the bottom curve is representative for this case.
A few remarks are in order. First, a significant gain
in the signal amplitude is evident for the wide aperture.
Second, the function $I_E(y)$ has support twice as wide as
the plasma diameter if $2\Delta=2R$ because for 
$R <|y|<2R$ no parallel ray comes through the plasma, but 
the aperture centered at such $y$ can still capture parallel
rays coming through the plasma. So the spatial resolution 
seems to be lost in the data $I_E(y)$. Third, 
the maximal amplitude of $I_E(y)$ cannot increase any further
if $2\Delta > 2R$. So, the aperture width equal to the plasma size
gives the maximal amplification of the signal.

To illustrate the restoration of the spatial resolution,
the function $I_E(y)$ is sampled at $y=y_n=
n\Delta_s$ where $\Delta_s = R/20$.
For the synthetic data $I_E(y_n)$ when $2\Delta = 2R$,
a discrete Fourier transform  is carried out
to compute $I'_E(y_n)$ as specified in (\ref{ff}). The Fast Fourier Transform Matlab package has been used for this purpose.
The values $I(y_n)$ are recovered by means of
the recurrence relation
(\ref{slit3}) with the initial conditions (\ref{slit4}).
They are shown by the (red) dots in
the middle panel of Fig.~\ref{fig2}. 
The dots lie on the graph of $I(y)$, i.e.,
the reconstruction 
is highly accurate.
Naturally, the accuracy of the emissivity reconstruction 
from the data $I(y_n)$ would then be fully determined by the 
accuracy of a particular Abel inversion algorithm \cite{abelinversion}.

\begin{figure}
 \centering
\includegraphics[height=6cm,width=17cm]{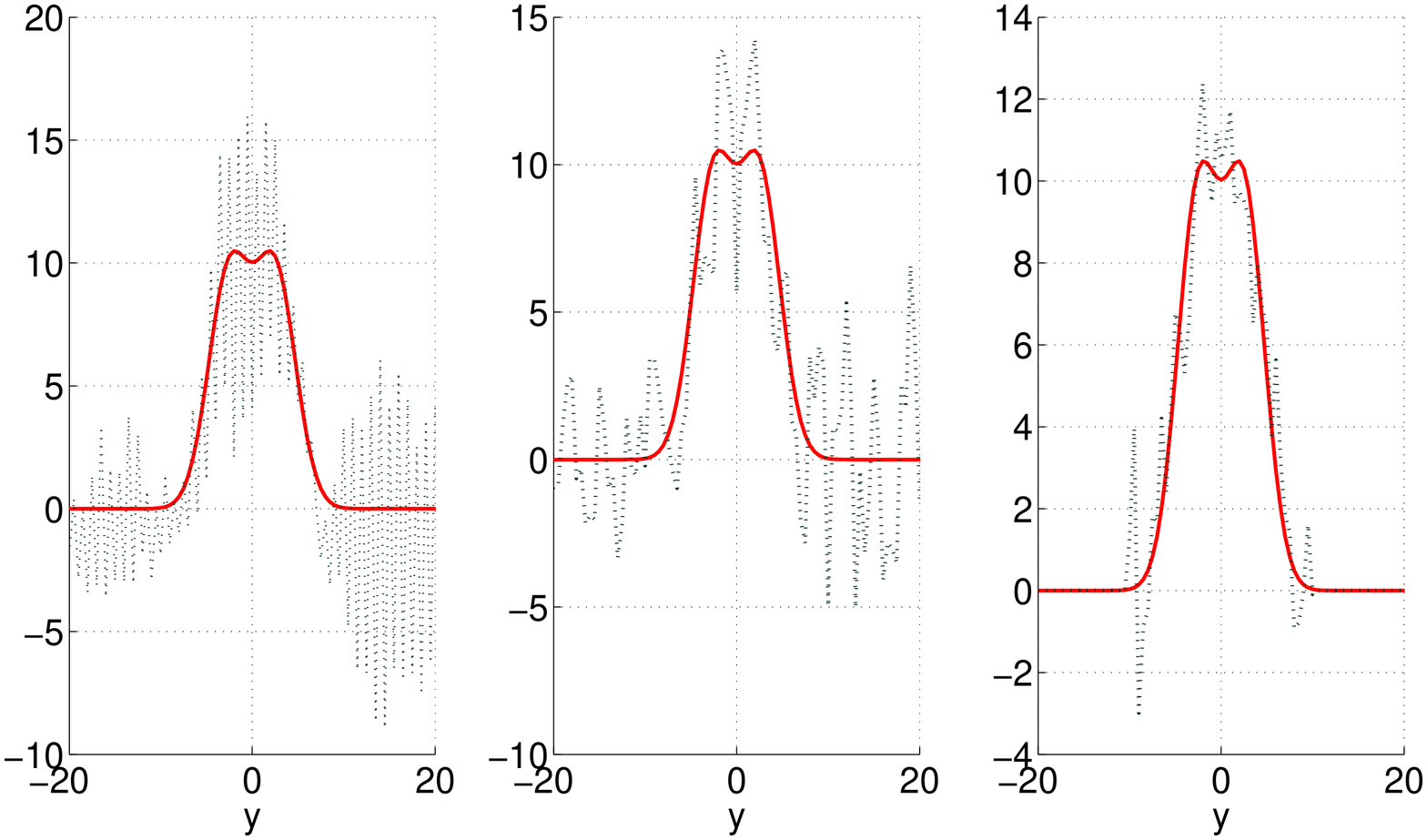}
\caption{\small Noise effects on the reconstruction algorithm.
The solid (red) curves on each panel is the graph of $I(y)$ as in
the middle panel of Fig.~\ref{2} but without normalization
on its maximal value. The dotted curves are the graphs 
of the reconstructed $I(y)$ from the "noisy" data $I_E(y)$
as explained in the text for $\Delta = R/10$ (left panel),
$\Delta =R/2$ (middle panel), and $\Delta =R$ (right panel).
The signal-to-noise ratio is $S=0.7$, $S=2.9$, and $S=3.1$,
respectively, in these three cases.}
\label{fig3}
\end{figure}
The proposed scheme 
can also be used to improve the signal to noise ratio when
the integration time is {\it fixed}.
In a typical LIBS experiment, the intensity data are collected
from repeatedly created plasma plumes. The power of a 
laser pulse  used for ablation varies from pulse to pulse,
a material surface on which the ablation process takes place
is not ideal, etc. All these effects generate a noise in $\ve(r)$
and, hence, in $I(y)$. As an example, a white noise 
$\eta(y)$ has been added to 
$I(y)\rightarrow I_\eta(y) =I(y)+\eta(y)$. The values of $I_E(y)$ have been calculated by 
Eq. (\ref{2}) with $I_\eta(y)$ for 50 samples of $\eta(y)$, and 
the result has been averaged. Then the reconstruction algorithm
has been applied to recover $I(y)$ from the "noisy" data $I_E(y)$.
Figure \ref{fig3} shows the graphs of the reconstructed functions
$I(y)$ (dotted curves) 
for $\Delta = R/10$ (left panel), $\Delta = R/2$ (middle panel), and 
$\Delta = R$ (right panel). The solid (red) curve in each panel is 
the actual intensity $I(y)$ corresponding to the emissivity
shown in the left panel of Fig.~\ref{fig2}. In contrast to 
$I(y)$ depicted in the middle panel of Fig.~\ref{fig2}, here
$I(y)$ is given without the normalization on its maximal value, i.e.,
as defined by Eq. (\ref{1}) without the factor $\Delta A\Delta\Omega$. 
The signal-to-noise ratio $S =I_m/\sigma$,
where $\sigma$ is the standard deviation of the white noise and
$I_m$ is the averaged signal amplitude near its maximum, is
$S=0.7$, $S=2.9$, and $S=3.1$ for the above three values 
of $\Delta$, respectively. The increase of the signal-to-noise
ratio in the reconstruction of $I(y)$ is self-evident.

\renewcommand\thesubsubsection{Acknowledgments}

\subsubsection{}

The authors acknowledge stimulating discussions with 
Prof. U. Panne (BAM) and D. Shelby (UF, Chemistry).
S.V.S. is grateful to Prof. U. Panne for
his continued support and thanks Department IV of BAM for a kind 
hospitality extended to him during his visit.
The work of I.B.G. is supported in part by 
the DFG-NSF grant GO 1848/1-1 (Germany) and 
NI 185/38-1 (USA).

\end{document}